\documentclass[aps,pre,showpacs,showkeys,preprint,floatfix,superscriptaddress]{revtex4-1}

\usepackage{amssymb}
\usepackage{amsmath}
\usepackage{amsfonts}
\usepackage{graphicx}
\usepackage{amsbsy}
\usepackage{color}
\usepackage{bm}
\usepackage{hyperref}

\newcommand{\bq}{\begin{eqnarray}}
\newcommand{\eq}{\end{eqnarray}}
\newcommand{\bqn}{\begin{eqnarray*}}
\newcommand{\eqn}{\end{eqnarray*}}

\newcommand{\rr}{\mathbf{r}}

\newcommand{\xx}{\mathbf{x}}
\newcommand{\yy}{\mathbf{y}}

\newcommand{\nablabf}{\bm{\nabla}}

\newcommand{\FF}{\mathbf{F}}
\newcommand{\SSS}{\mathbf{S}}

\begin{document}
\title{Form invariance of the moment sum-rules for jellium with
the addition of short-range terms in the pair-potential}

\author{Riccardo Fantoni}
\email{rfantoni@ts.infn.it}
\affiliation{Universit\`a di Trieste, Dipartimento di Fisica, strada
Costiera 11, 34151 Grignano (Trieste), Italy}

\date{\today}

\begin{abstract}
We find the first three (even) structure factor moments for a
(non-quantum) one-component Jellium made of particles living in three
dimensions and interacting with a Coulomb pair-potential plus a
short-range term with either a finite range or decaying exponentially
fast at large distances. Starting from the hierarchical
Born-Green-Yvon equations we show that they are all form
invariant respect to the addition of the short-range term. 
We discuss the relevance of the present study to
interpret the failure of the moment sum-rules of ionic-liquids at 
criticality.  
\end{abstract}

\maketitle
\section{Introduction}
\label{sec:intro}

A prototypical model of solid state physics describing free electrons
in metallic elements is the one-component {\sl Jellium}: a statistical
mechanics one-component fluid of point-wise charged particles made
thermodynamically stable by the addition of a uniform inert
neutralizing background. This fluid has been studied in great details
in history both in its classical and in its quantum versions. Here we
will only deal with the classical version of the model. In particular
several plausible exact relationships between the $n$-point
correlations functions, the so called {\sl sum-rules}, has been
determined over the years \cite{Martin1988}. Of particular interest,
due to the direct link with scattering experiments on the fluid, are
the even moments of the structure factor, the so called moment
sum-rules, which give the coefficients of the even powers of the wavenumber
in a large wavelength expansion of the structure factor. The
zeroth-moment sum-rule, or charge sum-rule, is commonly known as a
consequence of the {\sl internal} screening properties of the Coulomb
system and has been known since the work of Debye and H\"uckel
\cite{Debye1923}. The second-moment or Stillinger-Lovett
\cite{Stillinger1968} sum-rule is due to the {\sl external} screening
and has been proved rigorously for the first time by Martin et
al. \cite{Martin1983}. The fourth-moment sum-rule has been proved
rigorously for the first time by Vieillefosse \cite{Vieillefosse1985}
after it had been established earlier with various heuristic arguments
\cite{Pines,Vieillefosse1975,Baus1978}.  

A mixture of charged particles can have species with opposite
charges. In these cases in addition to the electrical neutrality of
the system with the introduction of a neutralizing background it is
necessary to introduce a hard-core on the particles, in order to
assure thermodynamical stability. It is then important to
understand how the addition of a short-range regularizing term (with
compact support or decaying exponentially fast \cite{Alastuey1985}) to
the pure Coulomb pair-potential influences the various sum-rules. 

In this work we perform this study on the one-component
Jellium extending Vieillefosse \cite{Vieillefosse1985} work to a
pair-potential where we add to the Coulomb term a generic short-range
term with either a finite support or exponentially decaying at large
distances. We work in three spatial dimensions leaving the extension 
to other dimensions, to a mixture, and to more general short-range
potential regularizations to future works. We start from the
constituent Born-Green-Yvon hierarchic equations
\cite{Hansen-McDonald} for the $n$-point correlation functions and
with certain assumptions on the decay of the $n$-particle Ursell
functions as subgroups of particles are infinitely separated (the {\sl
  exponential clustering} hypothesis) we use a series of {\sl
  multipolar sum-rules} \cite{Martin1988} to determine the first three
even structure factor moments.     
 
We already know that the forms of the first two even moments are not
influenced by the presence of the regularizing short-range term in the
pair-potential \cite{Martin1983,Martin1988}. We will find that also
the fourth-moment is form invariant.

This work is a step forward in the understanding of the failure of the
second and fourth moment sum-rules recently observed in the restricted
primitive model (RPM) at criticality \cite{Das2011,Das2012}. In fact,
it was only until recently that the previously unknown form of the
fourth moment sum-rule for the RPM was established using a semi-heuristic 
argument \cite{Fantoni16b,Fantoni17b} claiming form invariance respect
to the pure coulombic case. Our present result gives a rigorous first
principle proof of the form invariance at least in the weak coupling
regime and for the one-component Jellium. This fact, if confirmed for
the two-component plasma, relegates the failure of the sum-rules at
criticality to the disruption, upon approaching a phase transition, of
the {\sl exponential clustering}, i.e. the decay to zero of the
truncated correlations faster than any inverse power of distance as
groups of particles are separated by an infinite distance. In fact, it
has to be expected that at criticality the correlation functions
develops long-range tails with monotonous or oscillating inverse power
law decay \cite{Santos2015}. 

The work is organized as follows: In section \ref{sec:zeroth} we
find again the zeroth moment sum-rule following the original
derivation of Martin et al. \cite{Martin1988}, in section
\ref{sec:second} we find again the second moment sum-rule following
the original derivation of Martin et al. \cite{Martin1983,
  Martin1988}, in section \ref{sec:fourth} we derive the fourth moment
sum-rule with a route alternative to the one of Vieillefosse
\cite{Vieillefosse1985} which explains clearly from the point of view
of the BGY why also this moment is form invariant upon the addition of
a short-range term to the Coulomb pair-potential, in section
\ref{sec:compressibility} we determine the isothermal compressibility
of the system, and section \ref{sec:conclusions} is for the concluding
remarks.   

\section{Derivation of the zeroth moment sum-rule}
\label{sec:zeroth}
The second order Born-Green-Yvon (BGY) hierarchy \cite{Martin1988}
\bq \nonumber
\nablabf_{\rr_1}u_2(1,2)&=&\beta\FF_{21}[u_2(1,2)+1]+\\ \nonumber
&&\rho\int
d\rr_3\,[1+u_2(1,2)+u_2(1,3)+u_2(2,3)+u_3(1,2,3)]\beta\FF_{31}-\\ \nonumber
&&\rho\int d\rr_3\,[u_2(1,2)+1]\beta\FF_{31}\\ \label{BGY0}
&=&\beta\FF_{21}[u_2(1,2)+1]+\rho\int
d\rr_3\,[u_2(2,3)+u_3(1,2,3)]\beta\FF_{31}
\eq
where $\beta=1/k_BT$ with $k_B$ Boltzmann's constant and $T$ the
absolute temperature, $\rho$ is the density of the fluid,
$\FF_{21}=-\nablabf_{\rr_1}v(1,2)$, with $v(1,2)$ the 
pair-potential that is the sum of a Coulomb term
$v^c(1,2)=e^2/|\rr_2-\rr_1|$ and a short-range term $v^{sr}(1,2)$ with
compact support or decaying exponentially 
fast \cite{Alastuey1985}. We will also call
$\FF^c_{21}=-\nablabf_{\rr_1}v^c(1,2)$ and
$\FF^{sr}_{21}=-\nablabf_{\rr_1}v^{sr}(1,2)$. According to
Ref. \cite{Alastuey1985} the Ursell functions $u_n(1,2,\ldots,n)$ must
satisfy exponential clustering \cite{Martin1988}, i.e. they should
tend to zero (monotonously or oscillating) faster than any inverse
power of the distance as the distance between any group of particles at
$(\rr_1,\rr_2,\ldots,\rr_n)$ tends to infinity. The Ursell functions
are assumed to depend only on the shape of the figure formed by the
various points (and not on its space orientation) and they are
symmetrical in any permutation of the particles. The first assumption
is a consequence of the homogeneity and isotropy of the fluid, the
second is a consequence of distinguishability of the particles. Of
course the exponential clustering assumption is valid for the high
temperature (low density) infinite homogeneous phase of the fluid when
the correlation functions are believed to obey to the BGY
hierarchy. We will generally indicate vectors with a bold-face letter
and absolute values of vectors with a normal (Roman) version of the
same font $r=|\rr|$. We use a hat to denote the unit vector
$\hat{\rr}=\rr/r$. 

In the second equality of Eq. (\ref{BGY0}) we used the fact that $\int
d\rr_3\,u_2(1,3)\FF_{13}=0$ by symmetry. Now  
we observe that the left hand side of Eq. (\ref{BGY0}) tends to zero
faster than any inverse power of $x=|\xx|=|\rr_2-\rr_1|$ as $x$ tends
to infinity and the same is true for the first and fourth terms on the
right hand side. So the sum of the second and third terms on the right
hand side must vanish in the same way, in this limit. Then we require
that 
\bq \label{csr}
\int d\rr_3s(2,3)\FF_{31}
\eq 
where $s(2,3)=\rho u_2(2,3)+\delta(2,3)$ and $\delta$ is the Dirac
delta function, tends to zero faster than any
power of the distance $x$ when the latter tends to infinity. Expanding
Eq. (\ref{csr}) in powers of $1/x$ in this limit, we deduce
\bq \label{msr}
\int d\rr_3\, s(y)y^lP_l(\hat{\xx}\cdot\hat{\yy})&=&0~,~~~l\ge 1\\ \label{zmc}
I_0=\int d\rr_3\, s(y)&=&0
\eq 
where $\yy=\rr_3-\rr_2$ and $P_l$ are the Legendre
polynomials. Eq. (\ref{zmc}) is the zeroth moment sum-rule also known
as the charge or electroneutrality sum-rule \cite{Martin1988}. It is
the simpler of the multipolar sum-rules (\ref{msr}). We
immediately see that in our derivation we did not use the fact that
$v$ is purely Coulombic. It is sufficient that it contains the Coulomb
potential. 

\section{Derivation of the second moment sum-rule}
\label{sec:second}

Following Ref. \cite{Martin1983} we may write the second order BGY 
hierarchy as follows 
\bq \nonumber
\nablabf_{\rr_2}u_2(1,2)&=&\beta\FF_{12}[u_2(1,2)+1]+\\ \nonumber
&&\rho\int
d\rr_3[1+u_2(1,2)+u_2(1,3)+u_2(2,3)+u_3(1,2,3)]\beta\FF_{32}-\\ \label{BGY}
&&\rho\int d\rr_3[u_2(1,2)+1]\beta\FF_{32}
\eq
where $\FF_{12}=-\nablabf_{\rr_2}v(1,2)$, with $v(1,2)$ the
pair-potential, and the last line in Eq. (\ref{BGY}) is for the
neutralizing uniform background. We immediately observe that
$\int d\rr_3\,u_2(2,3)\FF_{23}=0$ by symmetry.

Multiplying by $\rr_{12}=\rr_1-\rr_2$ and integrating over $\rr_1$ we
find 
\bq \nonumber
\int d\rr_1\,\rr_{12}\cdot\nablabf_{\rr_2}u_2(1,2)&=&\int
d\rr_1\,\rr_{12}\bigg\{
\beta\FF_{12}+\\ \nonumber
&&\left.\rho\int 
d\rr_3\{u_2(1,3)+[\delta(1,3)+\delta(1,2)]u_2(2,3)/\rho+u_3(1,2,3)\}
\beta\FF_{32}\right\}\\ \nonumber
&=&\int d\rr_1\,\rr_{12}
\left\{\int d\rr_3[\rho u_2(1,3)+\delta(1,3)]\beta\FF_{32}+\right.\\
&&\left.\rho\int d\rr_3\,c_3(1|2,3)\beta\FF_{32}\right\},
\eq
where $s(1,3)=\rho u_2(1,3)+\delta(1,3)$ and
$c_3(1|2,3)=u_3(1,2,3)+[\delta(1,2)+\delta(1,3)]u_2(2,3)/\rho$ the
excess charge density which does not carry multipoles of any order
(See Proposition 2.2 in Ref. \cite{Martin1988}). Then  
\bq \nonumber
\int d\rr_1\,\rr_{12}\cdot\nablabf_{\rr_2}u_2(1,2)&=&
\int d\rr_1\rr_{12}\int d\rr_3s(1,3)\beta\FF_{32}+\\ \label{U3}
&&\rho\int d\rr_3\beta\FF_{32}\int d\rr_1\rr_{12}c_3(1|2,3).
\eq
Now we observe that due to the dipole sum-rule \cite{Martin1988} the
last line in Eq. (\ref{U3}) must vanish, $\FF=\FF^{sr}+\FF^c$ can be
split into a short-range term, $\FF^{sr}$, and a coulombic term,
$\FF^c$, where 
\bq
\int d\rr_1\,\rr_{12}\int d\rr_3\,s(1,3)\beta\FF^{sr}_{32}=
\int d\rr_{23}\,\beta\FF_{32}^{sr}\int d\rr_{13}\,(\rr_{13}+\rr_{32})s(1,3)=0,
\eq
where we used the charge sum-rule and isotropy of the system. This
tells us that the result we will find for the second moment is form
invariant under the addition to the pair-potential of a generic
short-range term. Also, using $\nablabf_{\rr_1}=-\nablabf_{\rr_2}$ and
$\int d\rr_1\ldots=-\int d\rr_2\ldots$, we
find   
\bq
\int d\rr_1\,\rr_{12}\cdot\nablabf_{\rr_2}u_2(1,2)&=&
3\int d\rr_1u_2(1,2)=-3/\rho,
\eq 
where we also used the charge sum-rule. Putting all together, we find 
\bq \nonumber
-\frac{3}{\rho}&=&\frac{1}{2}\int d\rr_1\nablabf_{\rr_1}(r_{12}^2)
\int d\rr_3 s(1,3)\beta\FF_{32}^c\\ \nonumber
&=&-\frac{1}{2}\int d\rr_1\nablabf_{\rr_2}(r_{12}^2)
\int d\rr_3 s(1,3)\beta\FF_{32}^c\\ \nonumber
&=&\frac{1}{2}\int d\rr_1 r_{12}^2\int 
d\rr_3 s(1,3)\beta\nablabf_{\rr_2}\FF_{32}^c\\
&=&\frac{1}{2}\int d\rr_1 r_{12}^2s(1,2)4\pi e^2\beta,
\eq
where we used the property that $\nablabf_{\rr_2}\FF_{32}^c=4\pi
e^2\delta(3,2)$. 
And finally we find for the second moment sum-rule
\bq \label{smc}
I_2=\int d\rr_2 r_{12}^2 s(1,2)=\frac{3}{2\pi\rho\beta e^2}=\frac{6}{k_D^2},
\eq
where $\lambda_D=k^{-1}_D=(4\pi\rho\beta e^2)^{-1/2}$ is the
Debye-H\"uckel screening length.

\section{Derivation of the fourth moment sum-rule}
\label{sec:fourth}

Starting from Eq. (\ref{BGY}) we multiply by $r_{12}^2\rr_{12}$ and
integrate over $\rr_1$ to get 
\bq \nonumber
\int d\rr_1\,r_{12}^2\rr_{12}&\cdot&\nablabf_{\rr_2}u_2(1,2)=\int
d\rr_1\,r_{12}^2\rr_{12}\bigg\{
\beta\FF_{12}+\\ \nonumber
&&\left.\rho\int
d\rr_3\{u_2(1,3)+[\delta(1,3)+\delta(1,2)]u_2(2,3)/\rho+
u_3(1,2,3)\}\beta\FF_{32}\right\}\\ \nonumber
&=&\int d\rr_1\,r_{12}^2\rr_{12}
\left\{\int d\rr_3[\rho u_2(1,3)+\delta(1,3)]\beta\FF_{32}+
\rho\int d\rr_3\,c_3(1|2,3)\beta\FF_{32}\right\}\\ \label{eq1}
&=&\int d\rr_1\,r_{12}^2\rr_{12}\int d\rr_3\,s(1,3)\beta\FF_{32}+
\rho\int d\rr_3\,\beta\FF_{32}\int d\rr_1\,r_{12}^2\rr_{12}
c_3(1|2,3).
\eq
Note that splitting again into a short-range term and the Coulomb one
we find for the first term on the right hand side of Eq. (\ref{eq1})
\bq \nonumber
\int d\rr_3\,\beta\FF_{32}^{sr}\int d\rr_1\,r_{12}^2\rr_{12}
s(1,3)&=&-\int d\rr_{32}\,\beta v^{sr}(3,2)\int d\rr_{12}\,
\nablabf_{\rr_{12}}[r_{12}^2\rr_{12}s(1,3)],\\ \label{eq2}
&=&\int d\rr_{32}\,\beta
v^{sr}(3,2)\int d\SSS_{12}\,r_{12}^2\rr_{12}s(1,3)=0. 
\eq
Since $s(1,3)=s(|\rr_{32}+\rr_{21}|)$ decays exponentially fast as
$r_{12}$ tends to infinity and the surface integral is over a sphere
centered on $r_{12}=0$ and with an infinite radius. The same holds for
the second term on the right hand side of Eq. (\ref{eq1}). 

This proves that the result we will find is independent from the
addition of a short-range part to the Coulomb pair-potential.

Now we observe that 
\bq
\int d\rr_1\,r_{12}^2\rr_{12}\int d\rr_3\,s(1,3)\beta\FF^c_{32}&=&
\frac{1}{4}\int d\rr_1\,\nablabf_{\rr_1}(r_{12}^4)
\int d\rr_3\,s(1,3)\beta\FF^c_{32}\\
&=&-\frac{1}{4}\int d\rr_1\,\nablabf_{\rr_2}(r_{12}^4)
\int d\rr_3\,s(1,3)\beta\FF^c_{32}\\
&=&\frac{1}{4}\int d\rr_1\,r_{12}^4
\int d\rr_3\,s(1,3)\beta\nablabf_{\rr_2}\FF^c_{32}\\ \label{eq3}
&=&\frac{1}{4}\int d\rr_1\,r_{12}^4s(1,2)4\pi e^2\beta.
\eq

And also using integration by parts
\bq 
\int d\rr_1\,r_{12}^2\rr_{12}\cdot\nablabf_{\rr_2}u_2(1,2)&=&\int
d\rr_1\,\nablabf_{\rr_{12}}(r_{12}^2\rr_{12})u_2(1,2)\\
&=&5\int d\rr_1\,r_{12}^2u_2(1,2)\\ \label{eq5}
&=&-\frac{15}{2\pi\rho^2\beta e^2},
\eq
where in the last equation we used the main result of previous section
for the second moment condition (\ref{smc}).

In this case $\int d\rr_1d\rr_3\,\beta\FF_{32}r_{12}^2\rr_{12}
c_3(1|2,3)\neq 0$ and we may recognize in such a term the one giving
rise to the isothermal compressibility in Eq. (\ref{fmc}).

Putting together Eqs. (\ref{eq1}), (\ref{eq2}), (\ref{eq3}),
(\ref{eq5}), and (\ref{chargesr}) we should reach the following fourth
moment result 
\bq \label{fmc} 
I_4=\int d\rr_2\,r_{12}^4s_2(1,2)=
\frac{15}{2\pi^2\rho^2\beta^2e^4}\frac{\chi_T^0}{\chi_T}=
\frac{120}{k_D^4}\frac{\chi_T^0}{\chi_T},
\eq
where $\frac{1}{\chi_T\rho}=\left.\frac{\partial P}{\partial\rho}\right|_T$ is
the isothermal compressibility and $\chi_T^0=\beta/\rho$ the one of
the ideal gas. As already stressed this result is independent from the
addition of a short-range term to the Coulomb pair-potential.

Then we should be able to prove that 
\bq\nonumber
1-\frac{\chi_T^0}{\chi_T}&=&
\frac{\rho\int d\rr_3\,\beta\FF^c_{32}\int d\rr_1\,r_{12}^2\rr_{12}c_3(1|2,3)}
{\int d\rr_1\,r_{12}^2\rr_{12}\cdot\nablabf_{\rr_2}u_2(1,2)}\\ \nonumber
&=&\frac{\rho}{5}
\frac{\int d\rr_3\,\beta\FF^c_{32}\int d\rr_1\,r_{12}^2\rr_{12}
c_3(1|2,3)}{\int d\rr_1\,r_{12}^2u_2(1,2)}\\ \nonumber 
&=&-\frac{2\pi\rho^3\beta^2 e^2}{15}
\int d\rr_{32}d\rr_{12}\,c_3(1|2,3)r_{12}^2\rr_{12}\cdot
\nablabf_{\rr_{32}}v^c(3,2)\\ \label{chargesr}
&=&-\frac{2\pi\rho^3\beta^2 e^2}{9}
\int d\rr_{32}d\rr_{12}\,c_3(1|2,3)r_{12}^2\rr_{32}\cdot
\nablabf_{\rr_{32}}v^c(3,2), 
\eq
where in the last equality we used $\rr_{12}=\rr_{13}+\rr_{32}$,
$\nablabf_{\rr}(r^2\rr)=5r^2$, $r^2\nablabf_{\rr}(\rr)=3r^2$, and
integration by parts. This will be done in the next section.  

\section{Compressibility sum-rule} 
\label{sec:compressibility}

From the virial theorem follows that the pressure estimator can be
written as follows \cite{Hansen-McDonald},
\bq
\beta P=\rho-\frac{\beta\rho^2}{6}\int
d\rr\,u_2(r)\rr\cdot\nablabf_{\rr} v^c(r).
\eq
So that 
\bq \label{compressibilitysr1}
1-\frac{\chi_T^0}{\chi_T}=
1-\beta\left.\frac{\partial P}{\partial\rho}\right|_T=\frac{\beta}{6}
\int d\rr&\,&\frac{\partial \rho^2u_2(r)}{\partial\rho}
\rr\cdot\nablabf_{\rr}v^c(r),
\eq
We then see that Eq. (\ref{chargesr}) can be obtained using an
analysis similar to the one of Vieillefosse et
al. \cite{Vieillefosse1989}, thus finding 
\bq \label{compressibilitysr2}
\frac{\partial \rho^2u_2(r_{32})}{\partial\rho}&=&
-\frac{4\pi\rho^3\beta e^2}{3}\int d\rr_{12}\,r_{12}^2c_3(1|2,3).
\eq

We then see how $\chi_T$ is the isothermal compressibility of a plasma
with a Coulomb interaction pair-potential among the particles.

\section{Conclusions} 
\label{sec:conclusions}

We determined the first three (even) structure factor moment sum-rules
(\ref{zmc}), (\ref{smc}), (\ref{fmc}) for a three-dimensional Jellium
with the particles interacting with a 
pair-potential that is the sum of the Coulomb potential $e^2/r$ and a
short-range term with either a finite range or decaying
exponentially fast at large $r$. We found that they are all invariant
in form respect to the addition of the short-range term. Moreover our
derivations of the sum-rules are different and simpler than the ones
already found in the literature (as described in the review of Ph. Martin
\cite{Martin1988}). This strategy carry us to the determination of an
compressibility sum-rule
(\ref{compressibilitysr1})-(\ref{compressibilitysr2}) in agreement
with the one of Vieillefosse \cite{Vieillefosse1989}. 
 
When studying common matter, whose constituents are made of charged
particles, the Coulomb interaction plays a special role, ruling the
fundamental correlation sum-rules. What really matter is the
long-range nature of the Coulomb interaction and the short-range
details do not have an influence on the statistical behaviors of the
many-body correlations. This allows to use different models for the
charges behavior at short-range where we may have some sort of
indeterminacy in the description of the point-wise constituents
particles microscopic character. All these models will have the same
macroscopic behavior.       

We could for example apply our general setting to the particular case
of charged hard-spheres, when the short-range term is just a hard-core
repulsion of a certain diameter. This is just one of the commonly used
short-range regularization employed in a two-component-plasma (TCP)
with particles of opposite charges
\cite{Fantoni16b,Fantoni13e,Fantoni13f} that would 
otherwise collapse one over the other. Moreover the hard-core model
has been historically the favorite playground in statistical mechanics
as it represents the simplest model of many-body systems of
interacting particles.  

In a recent work Das, Kim, and Fisher \cite{Das2011,Das2012} found out,
through finely discretized grand canonical Monte Carlo simulations,
that in the Restricted Primitive Model (RPM) of an electrolyte
\cite{Fantoni13e,Fantoni13f}, the second- and fourth-moment
charge-charge sum-rules, typical for ionic fluids, are violated at
criticality.   
For a 1:1 equisized charge-symmetric hard-sphere electrolyte their
grand canonical simulations, with a new finite-size scaling device,
confirm the Stillinger-Lovett second-moment sum-rule except, contrary
to current theory \cite{Stell1995}, for its failure at the critical
point $(T_c,\rho_c)$. Furthermore, the $k^4$ term in 
the charge-charge correlation or structure factor $S_{ZZ}(k)$
expansion is found to diverge like the compressibility when $T\to T_c$
at $\rho_c$. These findings are in evident disagreement with available
theory for {\sl charge-symmetric} models and, although their results are
qualitatively similar to behavior expected for {\sl charge-asymmetric}
systems \cite{Stell1995}, even a semi-quantitative understanding
has eluded them. Our present study could be a first step towards an
explanation of such puzzling behavior. Even if, as pointed out in
Ref. \cite{Fantoni17b}, from the work of Santos and Piasecki
\cite{Santos2015} follows that the Ursell functions of any order are
likely to have a long-range behavior on a critical point, thus
violating our exponential clustering working-hypothesis.

The zeroth-, second-, and fourth-moment sum-rules are rigorously
derived starting from the Born-Green-Yvon equations and the
exponential clustering hypothesis by Suttorp and van Wonderen
\cite{Suttorp1987,Wonderen1987,Suttorp2008} for a thermodynamically
stable ionic mixture made of point-wise particles of charges all of the
same sign immersed in a neutralizing background, the Jellium-mixture.
Our results show that the addition of a hard-core, or more generally
any finite-range or exponentially decaying contribution to the
pair-potential, to the particles, which would be necessary in order to
make thermodynamically stable the system of Suttorp and van Wonderen
for mixtures with particles of opposite charges, does not change the
form of the first two three moments of the structure factor of the
one-component Jellium. 

It is still an open problem the extension of our study to the more
general case of a mixture. A semi-heuristic derivation has
recently been carried out \cite{Fantoni16b,Fantoni17b} showing that the
addition of the short-range term should not play any role at the level
of the first three (even) structure factor moments for a neutral
TCP without the background. Strictly speaking, in these derivations we
had to use results that are only rigorously valid in the Debye regime,
like the local neutrality of the homogeneous system. 
Our present rigorous result confirms this scenario, at least in the
weak coupling limit. Another interesting project is to generalize
these sum-rule results to the case of Jellium living in curved surfaces
\cite{Fantoni03a,Fantoni08c,Fantoni12b,Fantoni12e}. In these cases the
system can be mapped in an equivalent flat Jellium interacting with an
external potential generated by the curvature of the surface in which
the particles live. Another interesting extension of our work consists
in studying the case in which the short-range pair-potential
decays at large distances as an inverse power $s$ of the distance, in
which case the decay of correlations is also always algebraic, with
the only exception of $s=\nu-2$ with $\nu$ the space
dimension \cite{Alastuey1985}. In this case we must drop the
exponential clustering hypothesis and our present derivation is not
valid anymore. 


\begin{acknowledgments}

\end{acknowledgments}

%

\end{document}